\documentclass{JFM-FLM_Au}

\newcommand{\bu}{\boldsymbol{u}}
\newcommand{\bv}{\boldsymbol{v}}
\newcommand{\be}{\boldsymbol{e}}
\newcommand{\dd}{\textrm{d}}
\newcommand{\buhat}{\boldsymbol{\hat{u}}}

\lefttitle{T.S. Eaves}
\righttitle{Journal of Fluid Mechanics}

\title{Minimal seeds in the Stokes boundary layer}

\author{Tom S. Eaves\aff{1}}

\affiliation{\aff{1} Division of Engineering, University of Dundee, Dundee DD1 4HN, UK}

\corresau{Tom S. Eaves, \email{teaves001@dundee.ac.uk}}

\begin{document}
	\maketitle
	
	\begin{abstract}
		Minimal seeds {\color{black} are} the smallest amplitude perturbations that trigger transition to turbulence {\color{black} in a subcritical flow. They} are presented in the Stokes boundary layer, the oscillating flow of a viscous fluid above a flat plate, {\color{black} at a fixed initial phase corresponding to the timing of the  Stokes boundary layer's optimal linear transient growth, at two moderate Reynolds numbers, and in two domains accommodating one and two wavelengths of the linearly optimal growing mode. Although} the minimal seed trajectories are dominated by the {\color{black} flow's} large linear transient growth at early times, only 77\% of the initial energy is formed from the linearly optimal growing mode; the remainder ensures that nonlinear interaction transfers energy from spanwise- to streamwise- independent structures, and makes up for a timing mismatch between the end of linear transient growth and the production phase of the edge state (the saddle point separating the laminar and turbulent basins of attraction).
	\end{abstract}
	
	
	\section{Introduction}
	
	The Stokes boundary layer is the flow of a sinusoidally oscillating viscous fluid above a flat plate, generated either by an oscillation of the plate itself with frequency $\omega$ and amplitude $U_0$, the wall-driven case, or by an oscillating pressure gradient driving a flow which oscillates with frequency $\omega$ and amplitude $U_0$ far above a stationary plate, the pressure-driven case. The {\color{black} equations} associated with each case are equivalent owing to a transformation between them, and they are linearly stable to normal mode perturbations growing over the whole cycle of oscillation \citep{blennerhassett2002linear} and to linear absolute convective instability \citep{ramage2020,pretty2021} below a critical Reynolds number of $\Rey_c \approx 2511$. Following \citet{biau2016transient}, the Reynolds number is given by $\Rey = U_0 \sqrt{T/\nu}$, where the oscillation period is $T = 2\upi/\omega$ and $\nu$ is the fluid viscosity (other definitions of the Reynolds number are a factor of $\sqrt{\upi}$ \citep[e.g.][]{kerczek74} or $2\sqrt{\upi}$ \citep[e.g.][]{blennerhassett2002linear} smaller than that used here). The critical Reynolds number is significantly larger than experimentally observed thresholds for transition \citep[see][]{blennerhassett2008linear}. While linear growth may occur instantaneously at smaller values of $\Rey$ \citep{kerczek74}, only finite-amplitude perturbations can trigger sustained turbulent episodes; the flow is subcritical. 
	
	The stability properties of subcritical flows are readily interpreted through the lens of dynamical systems. Both the laminar flow and turbulence are locally attracting within the state space, splitting it into two basins of attraction separated by an `edge' manifold. The question of nonlinear stability in these systems is therefore one of the shape and features of this edge. Recently, \citet{sandoval2025} described dynamics of the `edge state' (the saddle point separating laminar and turbulent trajectories) in the Stokes boundary layer, an irregular oscillation formed from cycles of streamwise streaks, rolls, and waves, a periodic version of the well-known self-sustaining process \citep{hall1991strongly,waleffe1997self}. While edge states provide insight into transitional dynamics, their properties are not correlated with nonlinear stability measures of the laminar flow \citep{pershin2022optimizing,eaves2025}. Instead, nonlinear stability is concerned with the energy of the smallest amplitude initial condition that triggers transition to turbulence (the so-called `minimal seed'). Minimal seeds may be identified via an optimisation process that finds initial conditions which grow substantially over long times whilst minimising their initial energy \citep{Kerswell2018}. The form and subsequent trajectory of the minimal seed represents the most efficient method of triggering turbulence in the flow, and the trajectory bootstraps distinct linear transient growth mechanisms together, and visits an edge state while doing so \citep{duguet2013}. In the Stokes boundary layer, \citet{biau2016transient} showed that optimal linear transient growth is very large (an $\mathcal{O}(10^6)$ increase in energy at $\Rey = 1000$), and that the optimally growing linear mode is a two-dimensional spanwise-independent structure which grows for a period of approximately $0.4T$ using the Orr mechanism \citep{orr}.
	
	In direct numerical simulations, \citet{ozdemir2014direct} observed intermittent turbulent episodes at $\Rey \approx 886$ and sustained (low-$\Rey$) turbulence at $\Rey \approx 1063$, {\color{black} both below the critical Reynolds number $\Rey_c$}. This observation is attributed to the nonlinear evolution of linearly unstable growing modes, and given that linear transient growth is very large in the system, it is likely that this transient growth is responsible for their conclusions. However, the flow is subcritical and transition must be an inherently nonlinear process. The optimally growing linear mode is two-dimensional, and so even if the minimal seed uses this linear growth during its trajectory, it must contain other three-dimensional structures that bootstrap on this growth to transition to turbulence. Indeed, the optimally growing linear mode is spanwise-independent, but the edge state is (nearly) streamwise-independent, and the minimal seed trajectory is expected to pass by this state. 
	
	\citet{Gong2022} present a sequence of events that may allow this transition from spanwise- to streamwise-independence and then to turbulence to occur: after reaching its maximum amplitude, the linearly optimal growing mode ejects a spanwise vortex from the boundary layer. Such vortices are unstable at finite amplitude, and their primary instability generates a streamwise flow close to the wall. This allows streamwise streaks to form via the lift-up mechanism \citep{liftup}, which are themselves ejected from the boundary layer, break down, and trigger bypass transition to turbulence. However, this description says nothing about the minimum initial amplitude for this to occur, and it fails to consider the relative timing between events. The latter point is highlighted by \citet{sandoval2025}, who show that the edge state dynamics (the near-wall streamwise streaks) consist of growth and decay phases which only allow lift-up driven growth within set time periods and spatial locations. The end of linear optimal growth corresponds with minimal production from the lift-up mechanism, and so the minimal seed trajectory must bridge this timing mismatch in order to transition between the linear mode and the edge state.
	
	The objective of this work is to compute minimal seeds in the Stokes boundary layer {\color{black} at a fixed initial phase corresponding to the timing of optimal linear transient growth}, in order to demonstrate exactly how their trajectories piece together the various structures available and the relative timing between them. Section 2 describes the methodology for computing minimal seeds. Section 3.1 describes the minimal seed at $\Rey = 1000$ for a {\color{black} pressure}-driven case in a domain matching the wavelength of the optimally growing linear mode of \citet{biau2016transient}. This is compared to a {\color{black} wall-driven} minimal seed in a larger domain, and a case with $\Rey=1200$ in section 3.2. Conclusions are drawn in section 4.
	
	\section{Methodology}
	
	The total velocity $\bu_{\textrm{tot}}$ can be decomposed into the laminar flow and departures from it, $\bu_{\textrm{tot}} = U(y,t)\be_x + \bu$, where $\be_x$ is the unit vector in the streamwise direction. The problem is nondimensionalised using the velocity magnitude $U_0$ and the time-scale $T$. The length-scale of the problem emerges diffusively and is given by the boundary layer thickness scale $\sqrt{\nu T}$. The pressure-scale is $\rho U_0^2 / \Rey$, where $\rho$ is the fluid density.  Within this scaling and decomposition, the Navier--Stokes equations read
	\begin{equation}
		\bu_t+ \Rey(\bu \bcdot \bnabla \bu + U \bu_x + v U_y \be_x) = -\bnabla p + \nabla^2 \bu, \quad \bnabla \bcdot \bu = 0,
	\end{equation}
	where $\nabla$ is the gradient operator, $v = \be_y \bcdot \bu$ is the wall-normal velocity, $p$ is the fluid pressure, and subscripts denote partial differentiation. 
	The laminar flow $U$ is given by $U_w$ and $U_p$ for wall-driven and pressure-driven cases respectively, where
	\begin{equation}
		U_w = \cos(2\upi t - \sqrt{\upi}y)\exp(-\sqrt{\upi}y), \quad
		U_p = \cos(2\upi t) - \cos(2\upi t-\sqrt{\upi}y) \exp(-\sqrt{\upi}y).
	\end{equation}
	{\color{black} In both cases, the flow admits the time-shift reflect symmetry $\mathcal{SR}:\, (t,u)\mapsto (t+1/2,-u)$, where $u=\be_x \bcdot \bu$ is the streamwise velocity, and so restriction to half a period is possible when discussing stability. Furthermore, since $U_p = \cos(2\pi t) - U_w$, there is a transformation which maps the flow with $U=U_w$ onto the flow with $U=U_p$, namely $\mathcal{T}:\, (x,u)\mapsto (-x + \Rey \sin(2\pi t) / 2\pi, -u)$. This transformation, together with the symmetry $\mathcal{SR}$, renders the two flows equivalent, with the same energy for $\bu$.}
	
	To search for minimal seeds, initial conditions which maximise the energy density $E$ after long times $\tau$ are found. Specifically, maxima are computed for the Lagrangian
	\begin{align}
		\mathcal{L} &= \frac{1}{2} \| \bu(\tau) \|^2 - \left[\bv,\bu_t+ \Rey(\bu \bcdot \bnabla \bu + U \bu_x + v U_y \be_x) +\bnabla p - \nabla^2 \bu \right] \nonumber \\ &- \left[ q, \bnabla \bcdot \bu \right] - \langle \bv_0, \bu(0)-\bu_0 \rangle - c\left(\| \bu_0 \|^2 - 2E_0 \right),
	\end{align}
	where $\bv$ is the adjoint velocity enforcing the momentum equation, $q$ is the adjoint pressure enforcing the continuity equation, and $\bv_0$ enforces the initial condition whose energy $E_0$ is enforced by $c$. Angled brackets represent inner products integrated over time, $[t_0,t_0+\tau]$, and square brackets represent inner products integrated over both space and time, $\Omega \times [t_0,t_0+\tau]$, where $\Omega$ is the fluid volume. The energy density is defined by
	\begin{equation}
		E \equiv \frac{1}{2} \| \bu \|^2 \equiv \frac{1}{2L_xL_z} \int_\Omega |\bu|^2 \,\, \dd \Omega.
	\end{equation}
	
	The variational problem $\delta \mathcal{L}=0$ is solved via `direct-adjoint-looping' to find the initial condition of energy $E_0$ that grows the most over time $\tau$, starting with $E_0$ well above the threshold for transition to turbulence, and lowering $E_0$ whenever an initial condition which triggers turbulence is found. This is continued until no further reductions in $E_0$ can be made whilst still triggering turbulence, and the residual $\mathcal{R} \equiv \| \delta \mathcal{L} / \delta \bu_0 \| ^2 / \| \bv_0 \|^2 $ is small \citep[typically $\mathcal{O}(10^{-3})$, see][]{rabin2012}. This approaches the minimal seed `from above' (the turbulent side; see \citet{eaves2025} for a detailed algorithmic description), {\color{black} and brackets the critical energy of the minimal seed, $E_c$, by $E_c^- < E_c < E_c^+$.}
	
	Minimal seeds are computed for the three sets of parameters listed in table \ref{table:params}, which include two domains, two Reynolds numbers, and wall- and pressure-driven Stokes boundary layers. {\color{black} Although the wall- and pressure-driven cases are equivalent, the equations for each case do differ, and so confidence in the results can be gained when a consistent set of dynamics arise from minimal seeds computed using the two different types of forcing.} Computations are performed using the DNS solver \verb|Diablo| \citep{taylor2008numerical}, using $N_x=64$, $N_y = 241$, and $N_z = 32$. The periodic streamwise $x$ and spanwise $z$ directions, {\color{black} of extent $L_x$ and $L_z$ respectively,} use Fourier modes, and the wall-normal $y$ direction uses second-order finite differences. The surfaces $y=0$ and $y=L_y$ are no-slip and stress-free respectively. {\color{black} Square domains with $L_x = L_z$ were used, with $L_x$ matching the streamwise extent of the optimally growing linear mode found by \citet{biau2016transient} at $\Rey = 1000$ (or twice this, for the case in a wider domain). The Reynolds number chosen for the `Baseline' results is close to the observed value for sustained turbulence \citep{ozdemir2014direct}.}
	
	A global minimal seed represents the smallest value of $E_0$ over all initial times, requiring an additional optimisation over $t_0$ which can be achieved via exhaustive search \citep{rabin2014} or incorporated into $\delta \mathcal{L}=0$ \citep{Kerswell2018}. However, owing to the large linear transient growth, the initial time was set at $t_0 = 0.0723$, the initial time of the optimal linear transient growth associated with the Baseline parameters of table \ref{table:params}. {\color{black} Although this precludes the computation of a fully global minimal seed across all possible phases,} it will be shown that this is a locally optimal choice for $t_0$ using numerical evidence that $\delta \mathcal{L} / \delta t_0 = 0$. The time horizon was set at $\tau = 3$ until this was insufficient to distinguish between laminar and turbulent trajectories near the minimal seed energy $E_c$; it was then increased to $\tau = 4$. The time-step was set at $10^{-5}$ to avoid divergence in the turbulent part of the trajectory, and was increased to $10^{-4}$ once trajectories close to $E_c$ spent minimal time near the turbulent attractor \citep[fixed time-steps simplify the checkpointing required to solve $\delta \mathcal{L}=0$, see, e.g., ][]{eavesthesis}.
	
	\begin{table}
		\begin{center}
			\def~{\hphantom{0}}
			\begin{tabular}{cccccccc}
				Label & $L_x$  & $L_y$   &   $L_z$ & $\Rey$ & Forcing & $E_c^-$ & $E_c^+$ \\[3pt]
				Baseline & 8.2   & 10.0    &  8.2 & 1000 & Pressure & $1.900 \times 10^{-9}$ & $1.925 \times 10^{-9}$  \\
				Wide & 16.4   & 10.0    &  16.4 & 1000 & Wall & $7.475\times 10^{-10}$ & $7.500 \times 10^{-10}$ \\
				High $\Rey$ & 8.2   & 10.0    &  8.2 & 1200 & Wall & $2.950 \times 10^{-11}$ & $2.975 \times 10^{-11}$ 
			\end{tabular}
			\caption{Parameters for the minimal seed {\color{black} calculations, with the label of each case, the streamwise domain length $L_x$, the wall-normal domain length $L_y$, the spanwise domain length $L_z$, the Reynolds number $\Rey$, the type of forcing, and lower ($E_c^-$) and upper ($E_c^+$) bounds on the energy $E_c$ of the corresponding minimal seed initial condition.}}
			\label{table:params}
		\end{center}
	\end{table}
	
	\section{Results}
	
	{\color{black} The upper ($E_c^+$) and lower ($E_c^-$) bounds on the} the critical energy of the minimal seeds are presented in table \ref{table:params}. As anticipated, $E_c$ is relatively small in all cases (compared to turbulence in this flow, which has $E_\textrm{turb}=\mathcal{O}(10^{-2})$). The ratio between $E_c$ and the edge state energy ($E_\textrm{edge}=\mathcal{O}(10^{-3})$, \citet{sandoval2025}) is also consistent with the $\mathcal{O}(10^6)$ linear growth at the Baseline parameters, and the $\mathcal{O}(10^8)$ linear growth at the High $\Rey$ conditions. Although the smallness and approximate scaling of $E_c$ is somewhat expected from the work of \citet{ozdemir2014direct} and \citet{biau2016transient}, these minimal seed energies nevertheless go some way towards interpreting the observations of the former, namely that the flow appears to be linearly unstable even though it is subcritical. At first glance, the critical energy of the Wide case appears to be significantly smaller than that for the Baseline case. However, $E$ is defined as an area average, and the domain area (in $x$ and $z$) is four times larger in the Wide case than in the Baseline case. The Baseline case actually represents a smaller total energy, since $4 E_{c,\textrm{Wide}} > E_{c,\textrm{Baseline}}$.
	
	\begin{figure}
		\centering
		\includegraphics[width=0.95\linewidth]{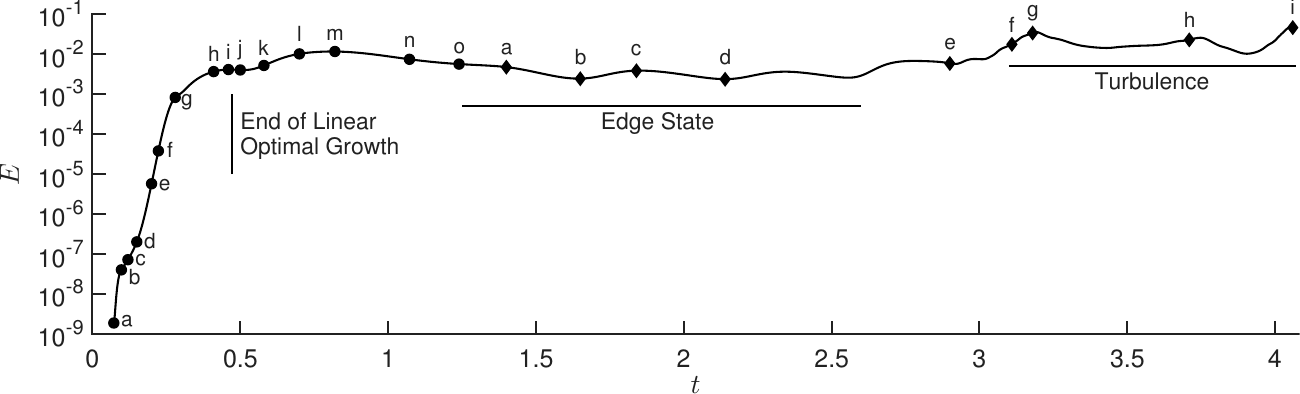}
		\caption{Energy $E(t)$ of the minimal seed {\color{black} trajectory} for the Baseline case. {\color{black} Circles and diamonds} indicate the times of the flow snapshots shown in figures \ref{fig:ms2edge} and \ref{fig:edge2turb} respectively, {\color{black} and the letter labelling identifies the associated panel within each of those figures}. {\color{black} Straight lines show the end of the linear optimal growth period at $t=0.4722$ identified by \citet{biau2016transient}, and approximate time intervals during which the trajectory is near the edge state and is in sustained turbulence.}}
		\label{fig:baseline_energy_plot}
	\end{figure}
	
	{\color{black} Focussing initially on only the Baseline case,} figure \ref{fig:baseline_energy_plot} shows the evolution of the energy $E(t)$, starting from the initial condition with $E=E_c^+$ at $t=t_0$, evolving up to the edge, and transitioning to turbulence. {\color{black} The initial growth is very rapid, up until the end of linear optimal growth at $t=0.4722$. There is then} a significant energy overshoot before edge state dynamics with $E_{\textrm{edge}} = \mathcal{O}(10^{-3})$ start around $t \approx 1.25$. {\color{black} The trajectory departs the edge state at around $t\approx 2.6$, gradually transitioning to sustained turbulence from $t\approx 3.1$. Circles and diamonds} in figure \ref{fig:baseline_energy_plot} indicate times of the flow snapshots shown in figures \ref{fig:ms2edge} and \ref{fig:edge2turb} respectively. Although the initial evolution of the minimal seed is dominated by linear growth, it must divide its initial energy between the linear optimal mode and some three-dimensional structures which grow alongside it, exchange energy with it through nonlinear interactions, and ultimately deposit energy into the structures associated with the edge state. The following section provides an in-depth description of these early-time dynamics {\color{black} and the transition process}.
	
	\subsection{Baseline case}
	
	\begin{figure}
		\centering
		\includegraphics[width=\linewidth]{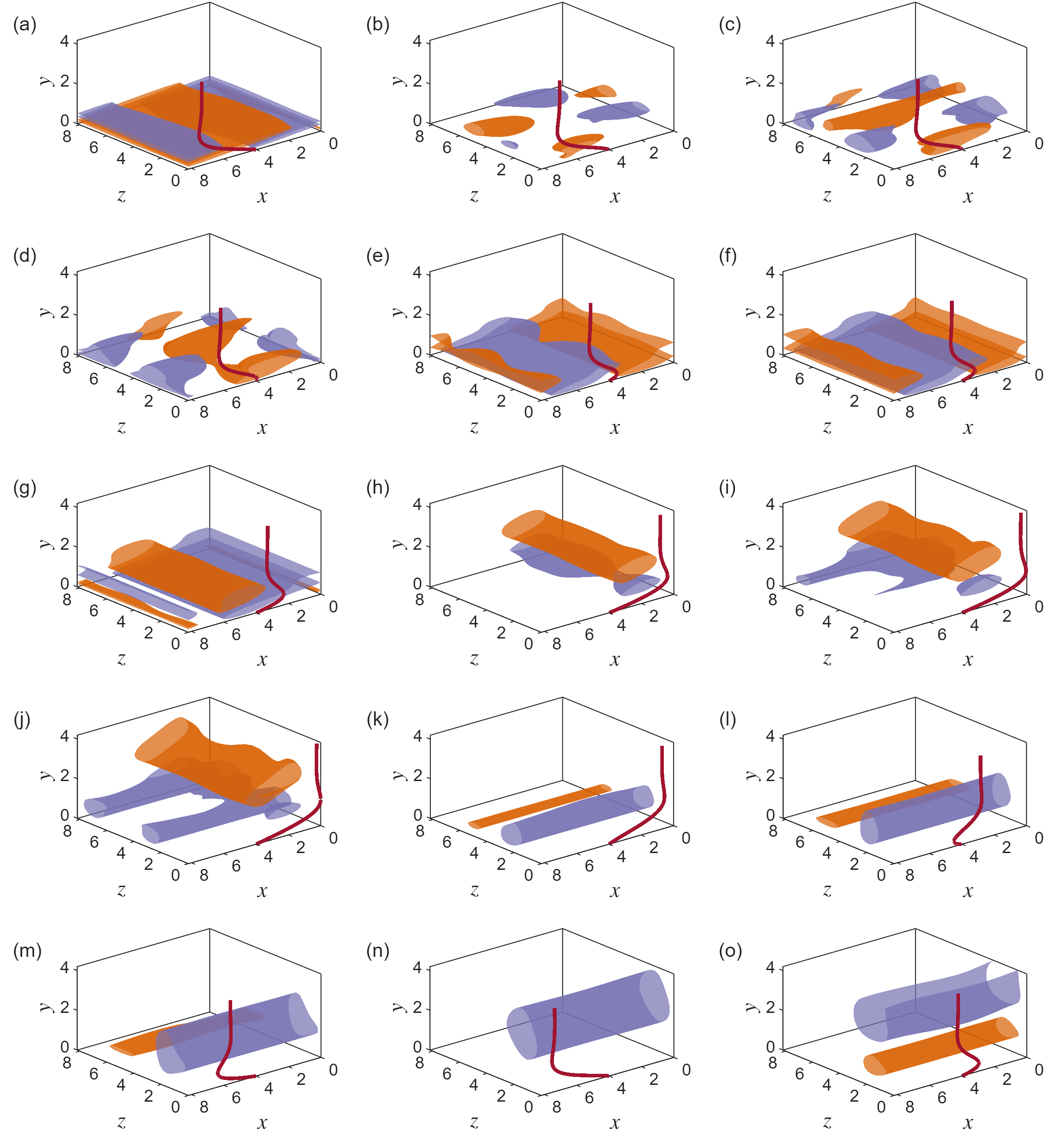}
		\caption{Isosurfaces of streamwise velocity $u= \pm 0.5 \max{|u|}$, with $u>0$ (blue) and $u<0$ (orange) at times {\color{black} indicated by black circles in figures \ref{fig:baseline_energy_plot} and \ref{fig:main_energy_plot}:} (a) $t = t_0 = 0.0723$, (b) 0.098, (c) 0.120, (d) 0.150, (e) 0.200, (f) 0.223, (g) 0.280, (h) 0.410, (i) 0.460, (j) 0.500, (k) 0.580, (l) 0.700, (m) 0.820, (n) 1.0723, and (o) 1.240. The full wall-normal extent $L_y = 10$ has been cut short to better show the details near the wall. The instantaneous laminar flow profile is indicated on the plane $z=0$ with a red line.}
		\label{fig:ms2edge}
	\end{figure}
	
	{\color{black} Figure \ref{fig:ms2edge} plots isosurfaces of streamwise velocity $u=\pm 0.5 \max{|u|}$ at the times indicated by black circles in figure \ref{fig:baseline_energy_plot}, spanning $t_0 \leq t \leq 1.24$. The initial condition (the minimal seed) consists of a finite-amplitude realisation of the spanwise-independent optimally growing linear mode, with a slight spanwise modulation (figure \ref{fig:ms2edge}(a)). This rapidly gives way to fully three-dimensional structures (figures \ref{fig:ms2edge}(b--d)), with evidence of streamwise-aligned structures briefly dominating the flow (figure \ref{fig:ms2edge}(c)). Spanwise-aligned structures then emerge once more (figure \ref{fig:ms2edge}(e)) and the flow evolves increasingly towards spanwise-independence (figures \ref{fig:ms2edge}(e--g)). This results in a spanwise-aligned structure being ejected from the near-wall region (figure \ref{fig:ms2edge}(h)), which is followed by the growth of streamwise streaks (figures \ref{fig:ms2edge}(i--j)) which ultimately dominate the flow (figures \ref{fig:ms2edge}(k--m)) and localise in the spanwise direction (figures \ref{fig:ms2edge}(n-o)). The emergence of three-dimensional and streamwise-aligned structures early in the linear optimal growth period stands in contrast to the transition sequence hypothesised by \citet{Gong2022}, in which the flow would remain spanwise-independent until it became linearly unstable as structures are ejected from the near-wall region.}
	
\begin{figure}
		\centering
		\includegraphics[width=0.95\linewidth]{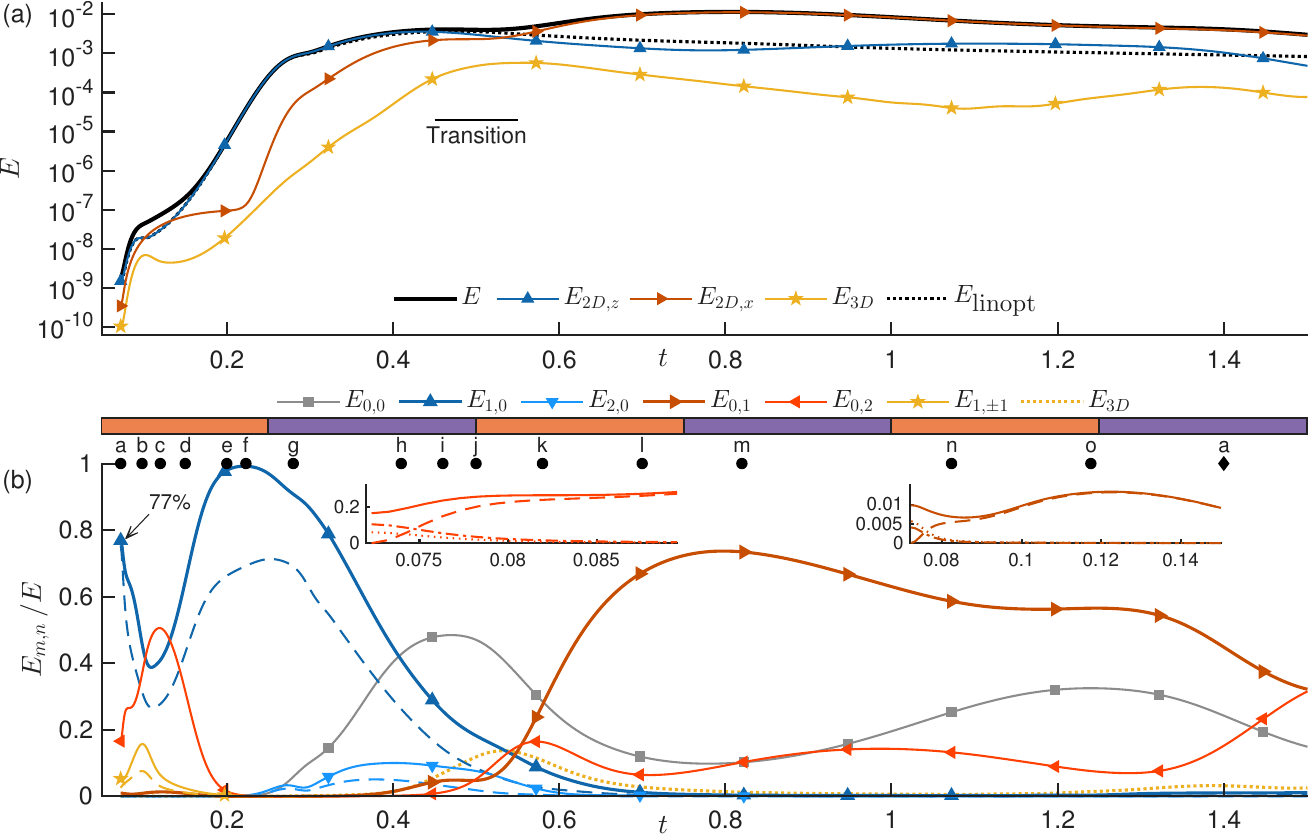}
		\caption{(a): Components of energy $E(t)$ of the Baseline minimal seed trajectory from its initial condition until its arrival at the edge state. Total energy (black thick), $E_{\textrm{2D},z}$ (blue upwards triangles), $E_{\textrm{2D},x}$ (red rightwards triangles), $E_{\textrm{3D}}$ (yellow stars), and $E_{\textrm{linopt}}$ (black dotted). {\color{black} The straight line shows the approximate period ($ 0.45 \lesssim t \lesssim 0.55$) during which the flow transitions from largely spanwise-independent to largely streamwise-independent modes.} (b): Ratios $E_{m,n}/E$ over the same period. {\color{black} Mean flow} $E_{0,0}$ (grey squares), {\color{black} spanwise-independent components} $E_{1,0}$ (dark blue upwards triangles) and $E_{2,0}$ (light blue downwards triangles), {\color{black} streamwise-independent components} $E_{0,1}$ (dark red rightwards triangles) and $E_{0,2}$ (light red leftwards triangles), and {\color{black} oblique combinations} $E_{1,\pm 1} \equiv E_{1,1}+E_{1,-1}$ (yellow stars) and $E_{3D}$ (yellow dotted). Dashed lines show the contribution to $E_{m,n}$ from the streamwise velocity alone if noticeably different from $E_{m,n}$. Insets show early-time breakdown of $E_{0,2}$ (left) and $E_{1,0}$ (right) with the contribution from the wall-normal component (dot-dashed) and the spanwise component (dotted). {\color{black} Black circles and diamonds along the top of panel (b) indicate the times of the flow snapshots shown in figures \ref{fig:ms2edge} and \ref{fig:edge2turb} respectively, and the letter labelling identifies the associated panel within each of those figures. The alternating orange-purple strip shows periods of far-field deceleration (orange) and acceleration (purple) according to the sign of $\dd |U_p| / \dd t$ as $y\rightarrow \infty$.}}
		\label{fig:main_energy_plot}
	\end{figure}

	{\color{black} To better quantify these observations, consider a spatial Fourier decomposition of the flow, to evaluate the energy contained within different spanwise, streamwise, and three-dimensional structures. Define Fourier modes $\buhat_{m,n}(y,t)$ via
		\begin{equation}
			\bu(x,y,z,t) \equiv \frac{1}{N_x N_z} \sum_{m=-N_x}^{N_x} \sum_{n=-N_z}^{N_z} \buhat_{m,n}(y,t) \exp(i(mkx+nlz)),
		\end{equation} 
		where $k=2\pi/L_x$ and $l=2\pi/L_z$. Note that $\buhat_{-m,-n}= \buhat_{m,n}^*$, where $^*$ denotes the complex-conjugate, given that $\bu$ is real. The energy associated with $\buhat_{m,n}$ is then given by
		\begin{equation}
			e_{m,n}(t) = \frac{1}{2(N_x N_z)^2} \int_0^{L_y} | \buhat_{m,n}|^2 \,\, \dd y,
		\end{equation}
		and $e_{-m,-n}=e_{m,n}$. This allows a restriction to consideration of modes $(0,n)$ for $0\leq n \leq N_z$ along with $(m,n)$ for $1\leq m\leq N_x$ and $-N_z \leq n \leq N_z$ by defining $E_{0,0}=e_{0,0}$ and $E_{m,n} = e_{m,n}+e_{-m,-n} = 2e_{m,n}$ for $(m,n) \neq (0,0)$.}
	With this convention, the total energy is given by $E=\sum_{m,n} E_{m,n}$. Define also the $z$-averaged {\color{black} (spanwise aligned)} energy in the two-dimensional $x$--$y$ plane to be $E_{\textrm{2D},z} = \sum_m E_{m,0}$ and the $x$-averaged {\color{black} (streamwise aligned)} energy in the two-dimensional $y$--$z$ plane to be $E_{\textrm{2D},x} = \sum_n E_{0,n}$ (and note that these two definitions share the mean flow energy $E_{0,0}$). Finally, define the fully three-dimensional energy to be $E_{\textrm{3D}} = \sum_{m\neq 0,\, n\neq 0} E_{m,n}$. Time series of $E$, $E_{\textrm{2D},z}$, $E_{\textrm{2D},x}$, and $E_{\textrm{3D}}$ are plotted in figure \ref{fig:main_energy_plot}(a) for $t_0\leq t\leq 1.5$. {\color{black} This figure also plots the energy of the two-dimensional linear optimal perturbation $E_{\textrm{linopt}}(t)$ of \citet{biau2016transient} which has been scaled so that $E_{\textrm{linopt}}(0)=E_{\textrm{2D},z}(0)$.}
	
	Until $t \approx 0.45$, $E_{\textrm{2D},z}$ and $E_{\textrm{linopt}}$ are essentially identical, showing that the large linear growth is unaffected by a finite amplitude realisation within the minimal seed trajectory, curtailed only slightly in its very final stage. However, {\color{black} apart from a very brief period immediately after $t=t_0$,} the linear optimal mode only dominates the total energy $E$ after $t \approx 0.14$. Instead, $E_{\textrm{2D},x}$ exceeds $E_{\textrm{2D},z}$ for $0.1 \lesssim t \lesssim 0.13$ during which period $E$ is significantly larger than $E_{\textrm{linopt}}$. The contribution from $E_{\textrm{2D},x}$ then stalls and is not a significant contributor to $E$ (independently of $E_{0,0}$) until $t\approx 0.55$, {\color{black} towards the end of} a transition from the linear optimal growth dynamics to the streamwise-independent modes required by the edge state. The purely three-dimensional term $E_{\textrm{3D}}$ is present throughout, though at a consistently lower level than the other terms. The pre-seeding of streamwise independent structures at significant amplitude within the initial condition {\color{black} and their dominance within the early dynamics} suggests that the transition onto the edge state is not solely due to the sudden emergence of such structures through an instability of the finite-amplitude linear optimal mode as suggested by \citet{Gong2022} (though this may still play a role).
	
	{\color{black} To more clearly demonstrate the behaviour of different modes within the overall dynamics,} figure \ref{fig:main_energy_plot}(b) shows the relative contribution to the total energy $E$ from each of the individual modes $(m,n)$ that contribute significantly to the dynamics. {\color{black} Specifically, these are the mean flow, $(0,0)$, the spanwise modes $(1,0)$ and $(2,0)$, the streamwise modes $(0,1)$ and $(0,2)$, and the oblique modes $(1,\pm 1)$, plotted in combination by defining $E_{1,\pm 1} \equiv E_{1,1} + E_{1,-1}$.} For most of these modes, the primary contribution to $E_{m,n}$ is from the streamwise velocity $u$, with a few exceptions also plotted within the figure. For those exceptions, the difference is made up by the wall-normal velocity $v$ {\color{black} except where detailed in an early-time inset}.
	
	The initial condition is formed mostly from the linear optimal mode $(1,0)$, the streamwise independent mode $(0,2)$, and the oblique modes $(1,\pm 1)$. {\color{black} Although the linear optimal mode $(1,0)$ is the largest contributor to the minimal seed initial condition, figure \ref{fig:main_energy_plot}(b) shows that it constitutes only around 77\% of the initial energy $E_c^+$.} The {\color{black} second largest mode (17\% of $E_c^+$) is the streamwise independent} mode $(0,2)$, formed from $v$ and $w = \be_z \bcdot \bu$, roll structures in the $y$--$z$ plane that rapidly decay whilst creating substantial streaks ($u$) via the lift-up mechanism {\color{black} (see the first inset in figure \ref{fig:main_energy_plot}(b))}. These streaks are responsible for increasing $E_{\textrm{2D},x}$ above $E_{\textrm{2D},z}$ {\color{black} during the early stages of growth}. Meanwhile, {\color{black} 5\% of $E_c^+$ is contained} in modes $(1,\pm 1)$, roughly evenly split between $u$ and $v$. Their energy grows rapidly via the Orr mechanism (given that both $u$ and $v$ grow in tandem). The combination and competition between these early-time dynamics {\color{black} results in the flow structures observed in} figures \ref{fig:ms2edge}(a--g). 
	
	{\color{black} From $t\approx 0.23$ both the spanwise harmonic mode $(2,0)$ and the mean flow $(0,0)$ begin to grow, the latter ultimately dominating the flow during much of the transition period from spanwise to streamwise aligned structures. The emergence of these modes} is associated with the {\color{black} formation and} ejection of a {\color{black} localised} spanwise vortex from the boundary layer and its destruction, shown in figures \ref{fig:ms2edge}(h--j). {\color{black} The energetic signature of this destruction is the growth of $E_{\textrm{3D}}$ from $t\approx 0.4$. Around 80\% of $E_{\textrm{3D}}$ is accounted for by modes $(1,\pm 1)$ and $(2,\pm 1)$ in this period, with the remainder in higher-order modes which rapidly dissipate.} After $t\approx 0.55$ this vortex effectively plays no role in the dynamics, as can be seed in figures \ref{fig:ms2edge}(k,l). Meanwhile, the streamwise-independent dynamics of the edge state establish. The mode $(0,1)$ is the only other in the initial condition with a meaningful amplitude (0.98\% of $E_c^+$, with the next highest being 0.01\%). Like mode $(0,2)$, it begins with energy in $v$ and $w$ which transfer energy to $u$ using the lift-up mechanism (see the second inset of figure \ref{fig:main_energy_plot}(b)), continuing to grow slowly in this way until $t\approx 0.45$. It is then held at near-constant amplitude for $0.45 \lesssim t \lesssim 0.51$ {\color{black} before growing rapidly and dominating the flow as it arrives at the edge state}.
	
	The sustaining of mode $(0,1)$ is the `delaying action' required to bridge between the end of the linear optimal growth ($t=0.4722$) and the streamwise-independent streaks of the edge state which cannot effectively grow during $0.30 \lesssim t \lesssim 0.55$ \citep[see e.g. figure 4b of][]{sandoval2025}. The edge state itself consists of only one sign of streak within each half-cycle, and so mode $(0,1)$ (which by itself would contain positive and negative streaks) requires the inclusion of modes $(0,0)$ and $(0,2)$. The former {\color{black} dominates the transition period} as discussed above, and the latter is present with significant amplitude in the initial condition. These three modes together result in the streak localisation that is evident through figures \ref{fig:ms2edge}(k--o).
	
	{\color{black} To quantify exactly how each of these modes grow, and identify the energetic transfers between them which allow for the transition from spanwise to streamwise independent dynamics,} figure \ref{fig:main_triads_plot} shows the instantaneous rate-of-change of $E_{m,n}$, {\color{black} along with contributions to $\dot{E}_{m,n} \equiv \dd E_{m,n}/\dd t$} from laminar flow production, dissipation, and triadic transfer from other modes. {\color{black} Specifically, the energy equation for mode $(m,n)$ reads
		\begin{equation}
			\dot{E}_{m,n} = P_{m,n} + D_{m,n} + \sum_{p,q} T_{p,q\rightarrow m,n},
		\end{equation}
		where the laminar flow production is given by
		\begin{equation}
			P_{m,n} = -\frac{\delta_{m,n}\Rey}{(N_x N_z)^2} \int_0^{L_y} \mathfrak{R}\left( \hat{u}_{m,n}^* \hat{v}_{m,n} \right) U_y \,\, \dd y, \quad 
			\delta_{m,n} \equiv \begin{cases}
				1  &  \textrm{if } (m,n)=(0,0) \\
				2  &  \textrm{if } (m,n)\neq(0,0)
			\end{cases},
		\end{equation}
		with $\hat{u}_{m,n}$ and $\hat{v}_{m,n}$ being the streamwise and wall-normal components of $\buhat_{m,n}$ respectively, $\mathfrak{R}(\cdot)$ denotes the real part, and $\delta_{m,n}$ accounts for the reality condition on the Fourier modes and the associated restriction to a smaller set of values for $m$ and $n$.
		The dissipation is given by
		\begin{equation}
			D_{m,n} = - \frac{\delta_{m,n}}{(N_x N_z)^2} \int_0^{L_y} \left( \left| \frac{\partial \buhat_{m,n}}{\partial y} \right|^2 + (m^2 k^2 + n^2 l^2 ) |\buhat_{m,n}|^2 \right) \,\, \dd y,
		\end{equation}
		and the triadic transfer $T_{p,q\rightarrow m,n}$ to mode $(m,n)$ involving mode $(p,q)$ is given in appendix A (and accounts for transfer involving both $\buhat_{p,q}$ and $\buhat_{-p,-q}$). This triadic transfer term includes modes $(m\pm p,n\pm q)$ which also participate in the exchange; the value of $T_{p,q\rightarrow m,n}$ is a measure of the rate of energy transfer to mode $(m,n)$ from triads involving mode $(p,q)$, and negative values imply that energy is being transferred from mode $(m,n)$ to other modes in the triads.}
	
	{\color{black} Firstly, the early-time insets in figures \ref{fig:main_triads_plot}(c,d,f) confirm that the initial growth of modes $(0,1)$, $(0,2)$ and $(1,\pm1)$ are due to the linear transient growth mechanisms asserted above (lift-up or Orr), given that the only contribution towards increasing $E_{m,n}$ arises from the laminar flow production $P_{m,n}$.
		
		Figures \ref{fig:main_triads_plot}(a,b,e) show that the growth of the harmonic mode $(2,0)$ from $t\approx 0.23$ is due to both laminar flow production and triadic transfer from the linear optimal mode $(1,0)$, whereas the growth of the mean flow $(0,0)$ at this time is from the linear optimal mode $(1,0)$ alone. However, shortly thereafter the harmonic mode $(2,0)$ also transfers a significant amount of energy to the mean flow $(0,0)$ so that the harmonic mode $(2,0)$ acts as a conduit for further energy to be transferred from $(1,0)$ to $(0,0)$. This, along with further transfer from $(2,0)$ to higher harmonics (not shown), leads to curtailed growth of mode $(2,0)$ relative to the mean flow $(0,0)$, which can be seen in figure \ref{fig:main_energy_plot}(b). The vortex ejection starting around $t\approx 0.4$ can be identified from figure \ref{fig:main_triads_plot}(f), which shows that the oblique modes $(1,\pm 1)$ receive energy from the linear optimal mode $(1,0)$ and, to a lesser extent, its harmonic $(2,0)$. While some of this energy is subsequently transferred to the mean flow $(0,0)$, much of it is either dissipated or transferred to higher-order modes within $E_{\textrm{3D}}$.
		
		\begin{figure}
			\centering
			\includegraphics[width=0.95\linewidth]{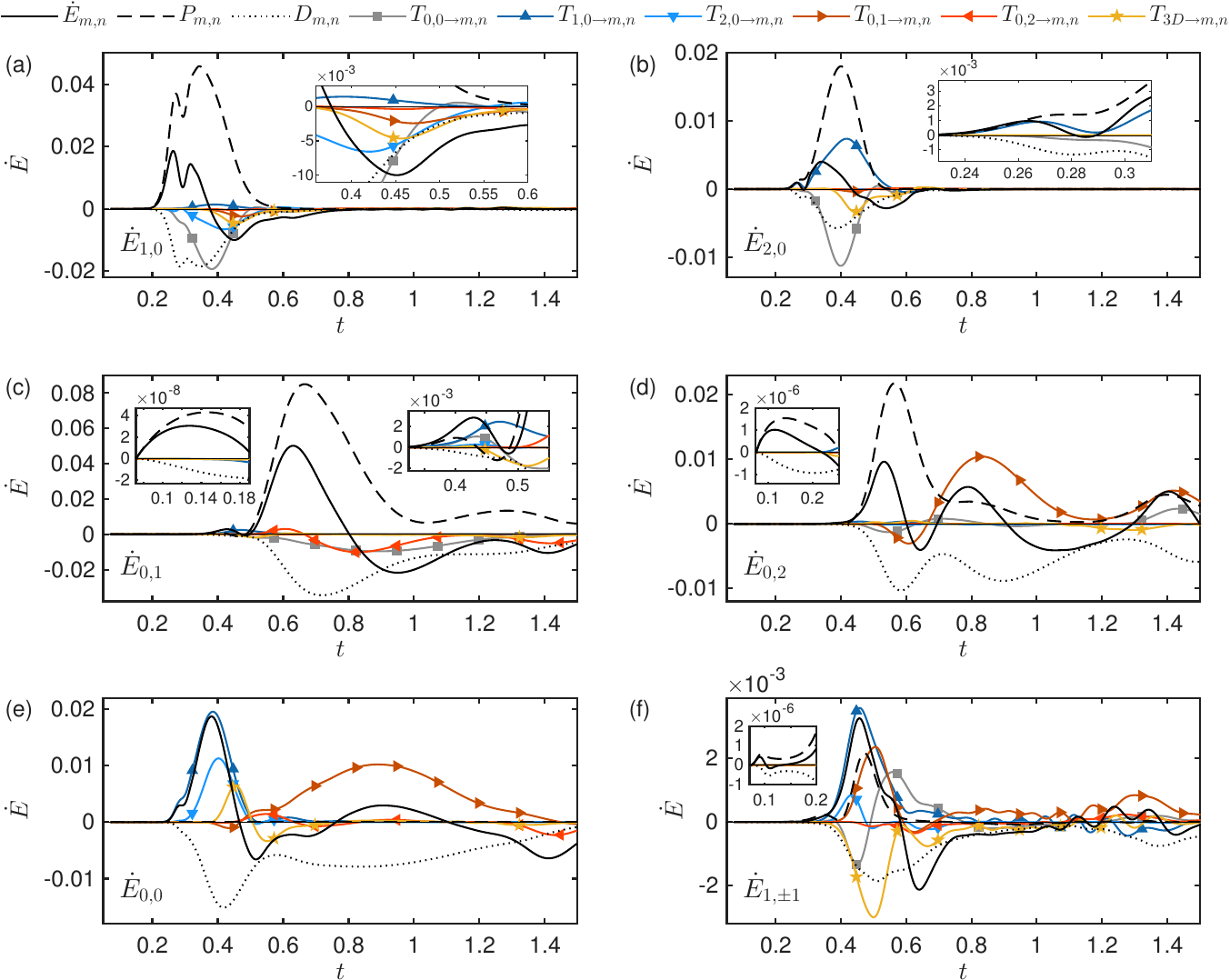}
			\caption{
				Energetic transfer to {\color{black} the spanwise-independent modes} (a) $E_{1,0}$ and (b) $E_{2,0}$, {\color{black} the streamwise-independent modes} (c) $E_{0,1}$ and (d) $E_{0,2}$, {\color{black} the mean flow} (e) $E_{0,0}$ and {\color{black} the oblique combination} (f) $E_{1,\pm 1} \equiv E_{1,1}+E_{1,-1}$. Total rate of change $\dot{E}_{m,n} \equiv \dd E_{m,n} / \dd t$ (thick black), laminar flow production (black dashed), dissipation (black dotted) and triadic wave transfers involving {\color{black} the mean flow} mode $(0,0)$ (grey squares), {\color{black} the spanwise-independent modes} $(1,0)$ (dark blue upwards triangles) and $(2,0)$ (light blue downwards triangles), {\color{black} the streamwise-independent modes} $(0,1)$ (dark red rightwards triangles) and $(0,2)$ (light red leftwards triangles), and {\color{black} the combination of oblique modes} $(1,\pm 1)$, $(1,\pm 2)$, $(2,\pm 1)$ and $(2,\pm 2)$ (yellow stars). {\color{black} Insets show detail at key moments discussed in the main text.}}
			\label{fig:main_triads_plot}
		\end{figure}
		
		This description provides the backdrop upon which to explain how the necessary sustaining `delaying action' of the streamwise-independent mode $(0,1)$ is organised during $0.45 \lesssim t \lesssim 0.51$. The second inset in figure \ref{fig:main_triads_plot}(c) shows that prior to $t\approx 0.4$ the mode $(0,1)$ is slowly growing due to laminar flow production (the lift-up mechanism), but that this production reverses sign to become laminar flow `destruction' during the holding period \citep[c.f.][]{sandoval2025}. Instead, it is} sustained by energy transfer from {\color{black} (in order of significance)} the {\color{black} the linear optimal mode} $(1,0)$, the mean flow $(0,0)$, and {\color{black} the harmonic mode} $(2,0)$ simultaneously. {\color{black} The transfer from the linear optimal mode $(1,0)$ is} a result of the spanwise instability of the ejected vortex, and is mediated by the oblique modes $(1,\pm 1)$ which were present in the initial condition with substantial amplitude. {\color{black} Transfer from the mean flow $(0,0)$ represents self-interaction amplification, and is only possible due to the creation of the mean flow described above. Finally, the small transfer from the harmonic mode $(2,0)$ is mediated by modes $(2,\pm 1)$ which were} created during the vortex destruction.

	{\color{black} Figure \ref{fig:main_triads_plot}(c) shows that after the holding period the streamwise-independent mode $(0,1)$ grows rapidly via laminar flow production. Streak localisation is initiated owing to the presence of the streamwise-independent harmonic mode $(0,2)$ in the initial condition, allowing for its own rapid growth via laminar flow production seen in figure \ref{fig:main_triads_plot}(d), and the creation of the mean flow $(0,0)$ by the linear optimal mode and its dynamics. From there, after a little transfer of energy from $(2,0)$ back to $(1,0)$, localisation is sustained by consistent energetic input to $(0,1)$ from laminar flow production and triadic transfer of energy from $(0,1)$ to both its harmonic $(0,2)$ and the mean flow $(0,0)$.}
	
	\begin{figure}
		\centering
		\includegraphics[width=\linewidth]{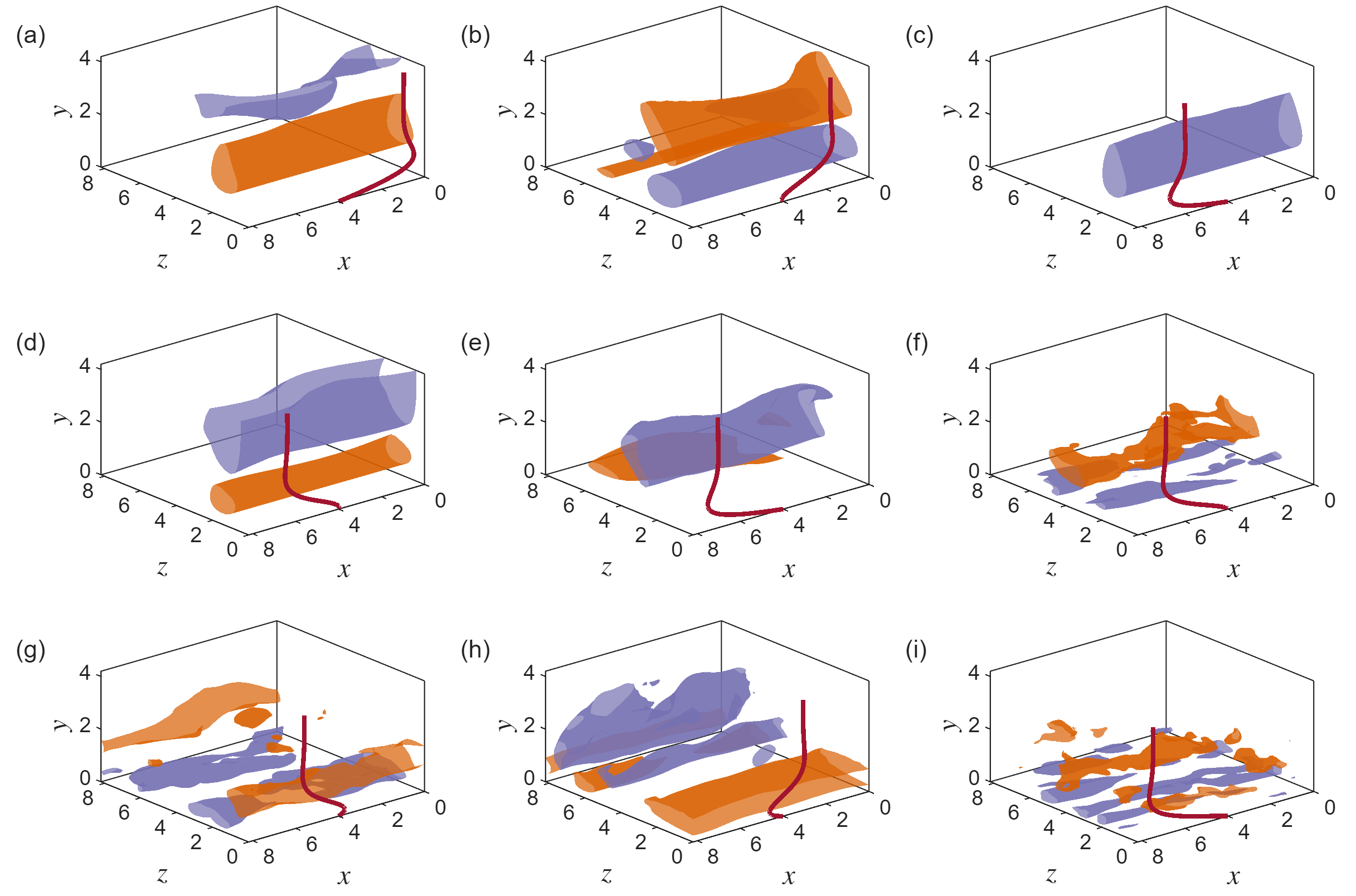}
		\caption{Same as figure \ref{fig:ms2edge} but for times {\color{black} indicated by black diamonds in figures \ref{fig:baseline_energy_plot} and \ref{fig:main_energy_plot}:} (a) $t=1.400$, (b) 1.650, (c) 1.840, (d) 2.140, (e) 2.900, (f) 3.110, (g) 3.180, (h) 3.710, and (i) 4.060.}
		\label{fig:edge2turb}
	\end{figure}
	
	{\color{black} To complete the description of the minimal seed's trajectory to turbulence, figure \ref{fig:edge2turb} plots isosurfaces of streamwise velocity $u=\pm 0.5 \max|u|$ at the times indicated by black diamonds in figures \ref{fig:baseline_energy_plot} and \ref{fig:main_energy_plot}, spanning the period $1.4<t<4.06$.} The edge state consists of alternatingly-signed streamwise streaks being ejected from the boundary layer (figures \ref{fig:edge2turb}(a--d); see \citet{sandoval2025} for a detailed analysis of this process). Eventually an ejection breaks down energetically enough to trigger a little turbulence in a similar way to bypass transition (figures \ref{fig:edge2turb}(e--f)). {\color{black} This turbulence spreads to occupy the spanwise extent of the domain (figures \ref{fig:edge2turb}(f--g)) and is then sustained (figures \ref{fig:edge2turb}(g--i)), a process common to other minimal seed trajectories \citep{duguet2013} albeit periodically modulated given that turbulence intensity is cyclical in the Stokes boundary layer, with peaks associated with the deceleration phase of the plate oscillation, and troughs being almost turbulence free at this moderate Reynolds number \citep{ozdemir2014direct}.}
	
	\begin{figure}
		\centering
		\includegraphics[width=\linewidth]{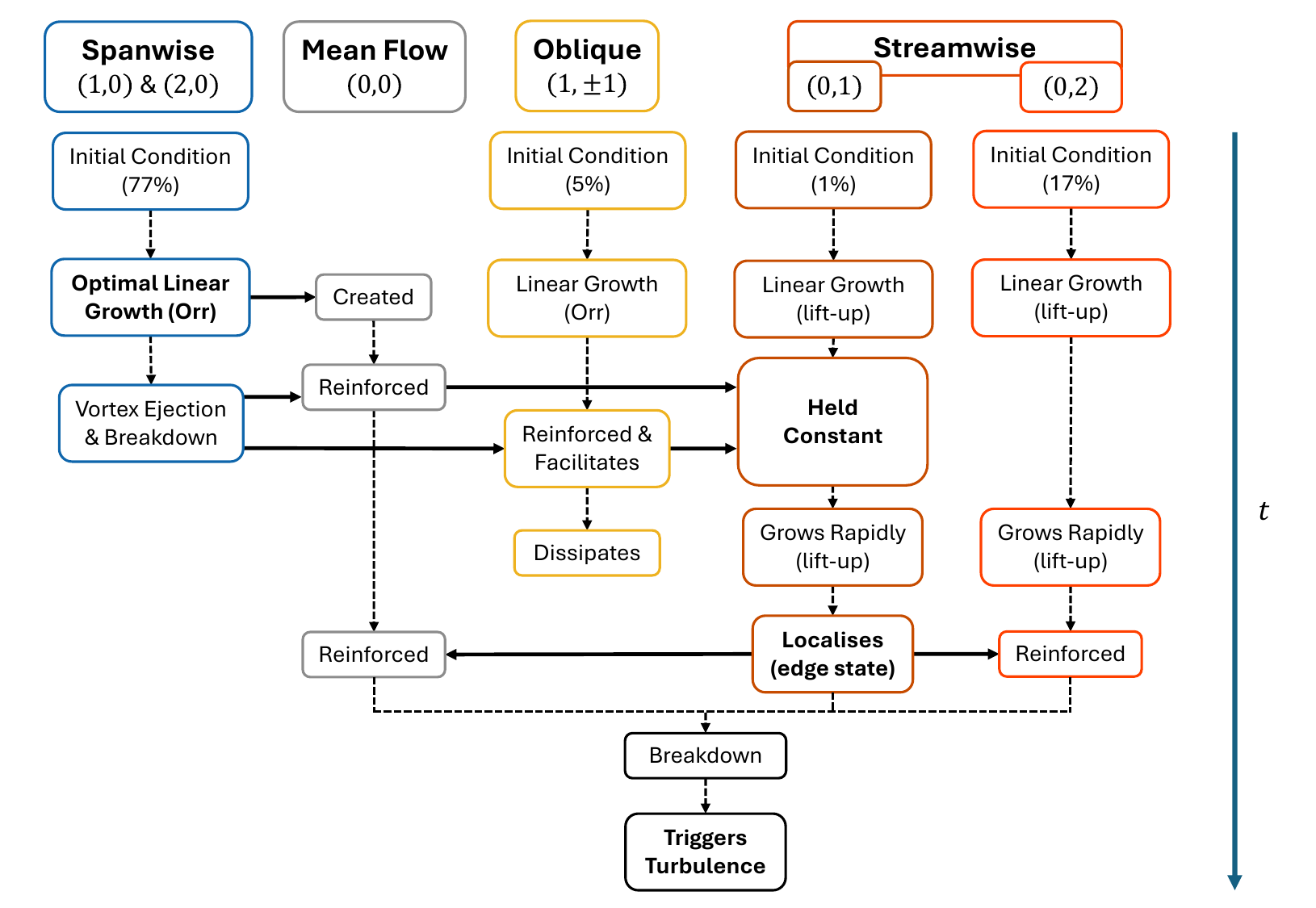}
		\caption{{\color{black} A diagram summarising the main features of the minimal seed transition route to turbulence, separated out into all spanwise-independent modes, the mean flow, oblique modes, and each streamwise-independent mode. Thick solid arrows indicate the main energy transfers between different modes.}}
		\label{fig:summary}
	\end{figure}
	
	{\color{black} Figure \ref{fig:summary} shows a diagram which summarises the minimal seed transition process.} The linear transient growth of $\mathcal{O}(10^6)$ is seemingly unavoidable {\color{black} at this choice of $t_0$} and the minimal seed makes use of this. {\color{black} The linear optimal mode is spanwise-independent and grows via the Orr mechanism. After saturating, it ejects a spanwise vortex from the near-wall region which breaks down and dissipates.} The edge state {\color{black} consists of streamwise-aligned streaks which} are orthogonal to the optimal linear growth mode, {\color{black} and so in order to make use of the large linear transient growth, oblique structures are present in the initial condition, grow via the Orr mechanism, and are further energised by the breakdown of the ejected vortex, in order to mediate the transfer of energy from the spanwise-independent linear optimal growth and the streamwise-aligned edge state.}  However, the edge state's growth periods are mistimed with respect to the end of the linear optimal growth phase, which concludes too early. This requires a sustained `holding' phase to allow for the timing mismatch, and is facilitated by {\color{black} continued transfer of energy from the linear optimal growth mode as it decays, alongside self-amplification arising from a substantial mean flow generated by the linear optimal growth. Once the streak growth period is reached, streamwise-independent structures grow rapidly due to the lift-up mechanism. The edge state consists of a single streak of one sign within each half-period of the wall oscillation, and so a single streamwise mode is insufficient (which would have one streak of each sign at all times). The initial condition therefore not only has} a very small component of the first streamwise-independent mode, {\color{black} but also} a substantial component of its first harmonic that grows via the lift-up mechanism throughout the linear transient growth phase. {\color{black} Together with the mean flow generated by the linear transient growth,} this allows spanwise localisation {\color{black} and a single sign of streak to develop.} {\color{black} Transition between the edge state and turbulence results from instability of these streaks, and the associated triggering of turbulent structures near the wall which eventually dominate the flow, particularly during the wall deceleration periods.} The dominant flow features align with those described by \citet{Gong2022}, but piecing them together into a single coherent trajectory proves particularly intricate.
	
	\subsection{Other cases}
	
\begin{figure}
		\centering
		\includegraphics[width=0.95\linewidth]{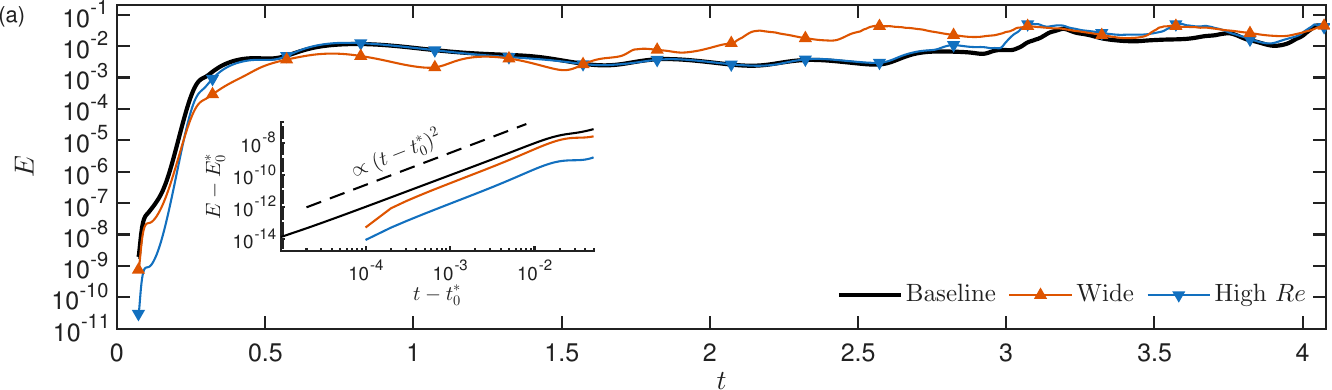}
		\includegraphics[width=0.95\linewidth]{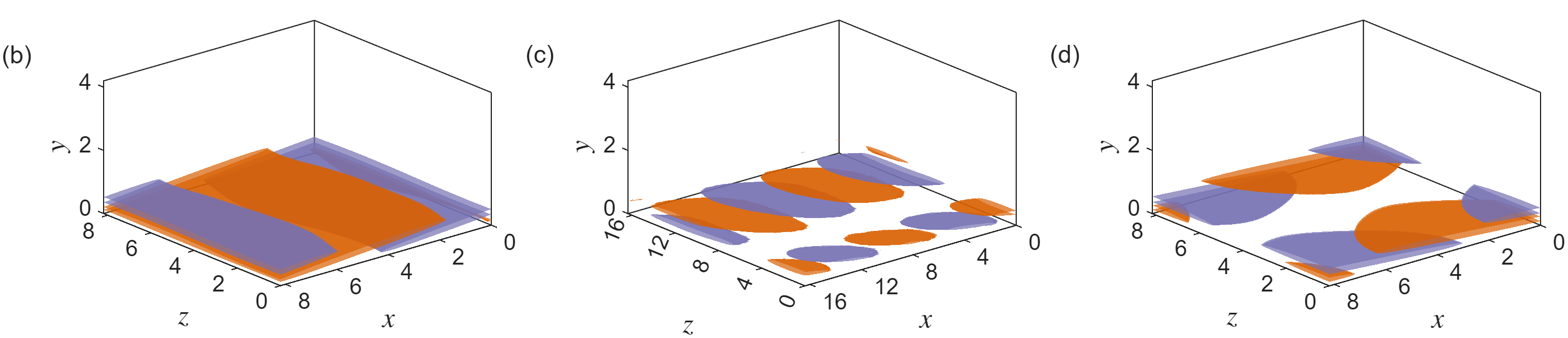}
		\includegraphics[width=0.95\linewidth]{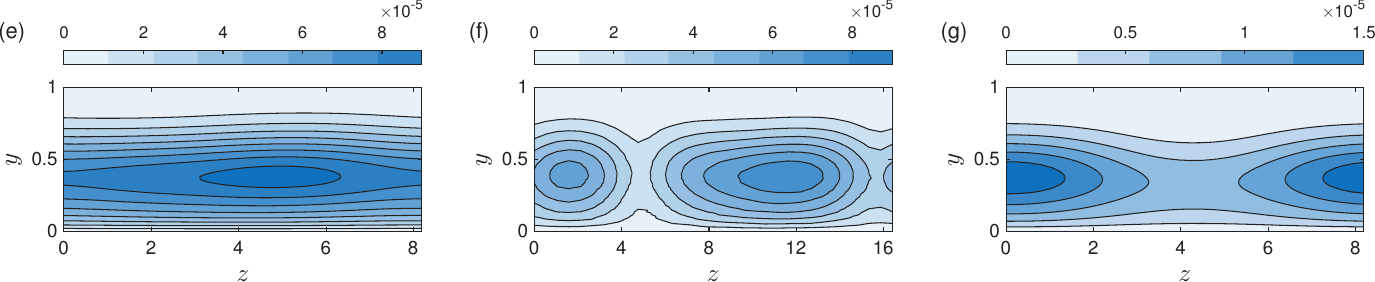}
		\caption{(a): Energy $E(t)$ of the minimal seed trajectories for the Baseline (black thick), Wide (red upwards triangles), and High $\Rey$ (blue downwards triangles) cases. Inset shows early-time {\color{black} scaling $E-E_0^* \propto (t-t_0^*)^2$, where $E_0^*$ and $t_0^*$ are the adjusted initial energy and initial time respectively, defined in the main text and given in table \ref{table:adjust}. (b-d): Isosurfaces of streamwise velocity $u=\pm 0.5 \max|u|$, with $u>0$ (blue) and $u<0$ (orange) for minimal seeds at $t=t_0$ in the (b) Baseline, (c) Wide, and (d) High $\Rey$ cases. (e-g): Streamwise average of the streamwise velocity magnitude $L_x^{-1}\int_0^{L_x} |u|\, \dd x$ at $t=t_0$ for the (e) Baseline, (f) Wide, and (g) High $\Rey$ cases.}}
		\label{fig:all_energy_plot}
	\end{figure}

	Figure \ref{fig:all_energy_plot}(a) shows the evolution of the total energy $E(t)$ for all three cases listed in table \ref{table:params}, starting from the initial condition with $E=E_c^+$ at $t=t_0$. Although the High $\Rey$ case starts with much smaller energy, from $t\approx 0.45$ its energetics are essentially identical to the Baseline case; once the linear growth phase is over the energetics  for $\Rey = 1200$ `catch up' and the subsequent nonlinear evolution is very similar. {\color{black} The Wide case is a little different to the others, in that it lacks a significant energy overshoot before reaching the edge. Such energy overshoots can be} common in other flows \citep[see e.g.][]{duguet2013,eaves2015disruption}, {\color{black} and the lack of an overshoot in the Wide case} will be explained below. {\color{black} The Wide case} also arrives at the edge state a half-cycle earlier than the other cases, {\color{black} and this will be linked to the lack of an overshoot. It also} leaves the edge state {\color{black} significantly earlier than the other cases}. However, {\color{black} that departure time} does not carry any significance {\color{black} in terms of the dynamics, but rather} reflects the closeness of $E_c^+$ to the true value of $E_c$ given that the edge state is a saddle in state space; {\color{black} the closer $E_c$ is to $E_c^+$, the longer the trajectory spends near this saddle, and indeed making the energy bracket $E_c^+-E_c^-$ arbitrarily small would result in an (inefficient) example of edge tracking with trajectories spending arbitrarily long periods of time near the edge state.}
	
	{\color{black} Figures \ref{fig:all_energy_plot}(b--d) show isosurfaces of streamwise velocity $u=\pm 0.5 \max|u|$ for each minimal seed at $t=t_0$.} Two movies of the trajectory of each minimal seed are available as Supplementary Data, {\color{black} with one movie showing the full trajectory from initial condition to turbulence, and another movie showing detail during the initial period $t_0<t<t_0+0.8$ by using a smaller frame-rate}. {\color{black} Figures \ref{fig:all_energy_plot}(c,d) show that there is a stronger spanwise modulation in the initial condition for the Wide and High $\Rey$ cases, with both} showing initial signs of spanwise localisation. {\color{black} This is confirmed in figures \ref{fig:all_energy_plot}(e--g) which plot the streamwise average of the streamwise velocity magnitude, $L_x^{-1} \int_0^{L_x} |u|\, \dd x$, for each minimal seed initial condition. It clear that spanwise localisation is incomplete in the Wide and High $\Rey$ cases.} Full localisation would require wider domains {\color{black} in each case}, and associated resolution requirements that place their computation beyond the feasibility of this current study. {\color{black} Since the linear optimal mode is spanwise-independent, it provides no emergent length-scale for any spanwise localisation. Instead, it is possible that the properties of the edge state itself set the required domain width. The edge state is necessarily localised so that only one sign of streak is present at a time. The spanwise extent of the edge state streak is given approximately by $\delta_{99}\approx 2.598$, where $\exp(-\sqrt{\pi}\delta_{99}) = 0.01$ so that $\delta_{99}$ is the 99\% boundary layer thickness. Smaller-amplitude structures surrounding the streak widen the edge state to around $2.5\delta_{99}$ in the domain of the Baseline case \citep[see figure 5 of][]{sandoval2025}. There is no reason to expect that the edge state becomes significantly wider as the domain width is increased given that the boundary layer thickness is the only length-scale inherent to the flow, and so complete localisation (with exponentially decaying tails either side of the edge state) might occur when the domain width is a few 10s of $\delta_{99}$.

		The evidence from the Baseline case suggests that} stronger spanwise localisation in the edge state would require additional streamwise-independent modes in the initial condition, {\color{black} and that the extended set of streamwise-independent modes needed to form the edge state will match those present in the initial condition. This provides a potential mechanism by which spanwise localisation of the minimal seed is achieved. Indeed, } the stronger spanwise localisation in the edge state of the Wide case requires additional modes in the initial condition, which {\color{black} leads directly to the prior observation that} $4E_{c,\textrm{Wide}} > E_{c,\textrm{Baseline}}$. The lack of an energy overshoot in {\color{black} the Wide case, and its earlier arrival at the edge state,} is due to an earlier spanwise localisation of the flow {\color{black} and a corresponding lack of a localisation phase after the end of linear transient growth. Except for this effect, the overall trajectory and dynamics of the Baseline and Wide cases are broadly similar.}
	
	The {\color{black} overall trajectory and dynamics of the} High $\Rey$ case {\color{black} are essentially identical to} the Baseline case, except that the linear transient growth of $\mathcal{O}(10^8)$ {\color{black} (compared to $\mathcal{O}(10^6)$ for the Baseline)} results in a value of $E_c$ about $10^2$-times smaller than the Baseline. The resources used to compute this `High' $\Rey$ case were substantial: approximately 307 days (24 hours per day) using 16 cores at 2.6 MHz ($\mathcal{O}(10^5)$ CPU-hours, albeit including inefficiencies associated with finding an appropriate time-step and the manual intervention between different values of $E_0$), a two-fold increase on the time taken for $\Rey = 1000$. For this reason, no further values of $\Rey$ were attempted (nor other values of $t_0$), but it {\color{black} could reasonably be hypothesised that if a wider range of Reynolds numbers are investigated,} the same essential understanding emerges, with increasingly large linear transient growth as $\Rey$ increases causing commensurate decreases in $E_c$.
	
	{\color{black} None of the minimal seeds show any sign of streamwise localisation. This is in contrast to minimal seeds in steady shear flows which typically localise in the streamwise direction at a similar length-scale to the spanwise direction \citep[see, e.g.][]{duguet2013}, even though their edge states require very large streamwise extents ($\mathcal{O}(\Rey)$) to localise \citep{khapko2016edge}. Although the Stokes layer advection time-scale ($\mathcal{O}(1/\Rey)$) is much shorter than the diffusive time-scale $(\mathcal{O}(1))$, it is possible that, unlike their steady shear flow counterparts, minimal seeds in the Stokes boundary layer have less time to be stretched into long streaks owing to the background shear reversing every half-period (and only being ``strong" for a yet shorter faction of the period). If this were indeed the case, it would imply that the streamwise extent of localised minimal seeds in the Stokes boundary layer should be a few times longer than for steady shear flow minimal seeds, to make up for this reduced stretching time.}
	
	\begin{table}
		\begin{center}
			\def~{\hphantom{0}}
			\begin{tabular}{cccc}
				Label & $E_0^* / E_c^+$  & $t_0^*-t_0$   &   $\Delta t$ \\[3pt]
				Baseline & 0.99998   & $2\Delta t$    &  $10^{-5}$ \\
				Wide & 0.99683   & $\Delta t$    &  $10^{-4}$ \\
				High $\Rey$ & 1.00000   & 0    &  $10^{-4}$
			\end{tabular}
			\caption{{\color{black} Adjusted initial energy $E_0^*$ and initial time $t_0^*$ for each minimal seed.}}
			\label{table:adjust}
		\end{center}
	\end{table}	
	
	{\color{black} The minimal seeds presented herein are all at fixed initial phase $t_0 = 0.0723$, the initial time of the optimal linear growth for $\Rey = 1000$. This precludes their designation as globally optimal initial conditions. Nevertheless, this choice of initial phase can be shown to be (approximately) locally optimal.} A necessary condition for the {\color{black} optimal} timing of a minimal seed is $\dd E / \dd t =0$ at $t=t_0$ \citep{Kerswell2018}. {\color{black} For most of the minimal seeds presented here, the time $t=t_0$ is not an exact minimum of $E(t)$, with $E(t)$ actually decreasing a small amount over the first few timesteps, but it is very close to being so. Define an adjusted initial energy $E_0^*$ and initial time $t_0^*$ by $\dd E / \dd t =0$ at $t=t_0^*$ and $E_0^* = E(t_0^*)$. The values of both $E_0^*$ and $t_0^*$ are given in table \ref{table:adjust} for all cases, where it is seen that there is very little difference between $t_0$ and $t_0^*$, differing by at most one or two computational timesteps.} The first inset of figure \ref{fig:all_energy_plot}(a) shows that {\color{black} $E-E_0^* \propto (t-t_0^*)^2$ in all cases,} i.e. that $E(t)$ is locally quadratic with a minimum at $t=t_0^*$. {\color{black} Therefore, to a very close approximation, the initial time $t=t_0$ represents a locally optimal initial phase.} Although there could be other {\color{black} initial times} which are more optimal seeds for turbulence {\color{black} (with lower values of $E_c$), they would need to bootstrap together several linear growth mechanisms before reaching the edge state in order to} beat the large {\color{black} optimal} linear growth.
	
	{\color{black} Finally, a comment is made on the robustness of the computed minimal seeds. Several versions of the Baseline case were computed, including wall- and pressure-driven cases, initialising from both random noise only and the linear optimal mode with added random noise (where the random noise was scaled to decay exponentially away from the wall with a decay length-scale matching the background flow). All cases approached the same set of dynamics and energetics described in section 3.1 (indeed, an earlier version of this work presented a wall-driven case for the Baseline). The wall-driven cases were not taken all the way to convergence since they stalled for an extended period of time with residual $\mathcal{R} = \mathcal{O}(10^{-2})$ and adjusted time $t_0^* \approx 45 \Delta t$ and it would have taken prohibitively long to reduce $\mathcal{R}$ to $\mathcal{O}(10^{-3})$ and $t_0^*$ to a small multiple of $\Delta t$. Nevertheless, their overall behaviour, and the consistency between these and the other cases described herein, lend some confidence towards the robustness of the results.}
	
	\section{Conclusions}
	
	Minimal seeds were computed in the Stokes boundary layer {\color{black} at fixed initial phase corresponding to the timing of optimal linear transient growth}, for two domain sizes and Reynolds numbers. In all cases, the minimal seed trajectory transitions between the structures posited by \citet{Gong2022}, that of the linear optimal transient growth mode expelling a spanwise vortex from the boundary layer, whose instability generates streamwise-aligned streaks which grow and rise up through the boundary layer, eventually triggering turbulence as they break down. However, the interaction between these components is surprisingly intricate, with the three-dimensional form of the initial condition allowing for a mismatch in timing between the linear transient growth and the dynamics of the edge state, and its spanwise localisation \citep{sandoval2025}. {\color{black} The requirement for the minimal seed trajectory to account for the timing of the edge state sets it apart from minimal seeds in steady shear flows.}
	
	{\color{black} For the main set of parameters considered here, only 77\% of the initial condition is formed from the linearly optimal growing mode. Since that mode is spanwise-independent, there must be some additional structure in the initial condition which allows for the streamwise-aligned structure of the edge state to form. A small amount (1\%) of the dominant edge state mode is present in the initial condition, along with oblique modes (5\%) that are required to facilitate triadic transfer of energy from the linear optimal mode to the dominant edge state mode once the former saturates. Finally, the initial condition has a large component of the dominant edge state mode's first harmonic (17\%) so that the edge state can localise, a requirement in the Stokes layer since the edge state is necessarily formed from only one sign of streamwise streak during each half-period of oscillation \citep{sandoval2025}. All of these initial modes grow initially using linear transient growth, arising from the Orr mechanism for modes with a spanwise component, and the lift-up mechanism for all streamwise-independent modes.
		
		As the optimal linear growth mode saturates and a spanwise vortex is ejected away from the wall, it not only transfers energy directly to the dominant edge state mode via triadic interaction with the oblique structures, but it also creates a substantial mean flow. This mean flow allows for self-reinforcement of the dominant edge state mode, and together with the direct triadic transfer holds the dominant edge state mode at constant amplitude until the timing is right for it to grow. Once the edge state growth window is reached, both the dominant edge state mode and its harmonic grow rapidly via the lift-up mechanism, and the large mean flow is sustained by transfer back from the dominant edge state mode. Together, these streamwise-independent modes create the necessary localisation for the edge state to exist. Eventually the balance associated with the edge state is lost, with a streak breaking down energetically enough to trigger turbulence near the wall.}
	
	{\color{black} It was shown that} the linear optimal growth initial time provides a locally optimal initial time for the minimal seed. It is likely {\color{black} on physical grounds, though not proven,} that this is globally optimal owing to the large linear transient growth present in the problem. This growth, and the minimal seed's utilisation of it, helps to explain the observations of \cite{ozdemir2014direct} that turbulence is almost unavoidable in the problem, and the subsequent hypothesis that the flow is in some way (linearly) unstable. 
	{\color{black} When the Reynolds number was increased so that the linear transient growth increased by $\mathcal{O}(10^2)$, the critical energy of the minimal seed reduced correspondingly by a factor of $\mathcal{O}(10^2)$. If the globally optimal initial time can be shown to be set by the linear optimal growth, then it should be possible to associate properties of this growth to the scaling of the critical energy of the minimal seed with the Reynolds number.} 
	
	{\color{black} Despite doubling the domain size, a fully localised minimal seed was not found, although increased spanwise modulation was observed. Hypotheses were offered which suggest that the spanwise extent required for localisation of the minimal seed should be linked to the size of the edge state, and that the apparent lack of streamwise modulation might be due to a shorter time available for structures to stretch in the streamwise direction owing to the oscillating shear.} Thorough analyses of the globally optimal initial time, clear demonstration of a scaling with Reynolds number, and identification of fully localised minimal seeds in large domains are warranted, albeit tasks that are well beyond the resources available to this work. {\color{black} Such an investigation may uncover less obvious nonlinear transition routes which do not rely on the optimal linear transient growth.}
	
	\begin{bmhead}[Supplementary data.]\label{SupMat}
		Movie versions of figures \ref{fig:ms2edge} and \ref{fig:edge2turb} are available for all four parameter values.
	\end{bmhead}
	\begin{bmhead}[Acknowledgements.]
		Prof. R. Kerswell and Mr. R. Dahan are thanked for discussions about their work. Three anonymous reviewers are thanked for their insights which helped improve the manuscript.
	\end{bmhead}
	\begin{bmhead}[Funding.] 
		This research was funded by an EPSRC New Investigator Award (EP/W021099/1).
	\end{bmhead}
	\begin{bmhead}[Declaration of interests.]	
		The author reports no conflict of interest.
	\end{bmhead}
	\begin{bmhead}[Data availability statement.]
		The initial condition and grid file associated with each minimal seed are available at \url{https://doi.org/10.15132/10000301}.
	\end{bmhead}
	\begin{bmhead}[Author ORCID]
		T. Eaves, \url{https://orcid.org/0000-0003-3473-1306}
	\end{bmhead}			
	
	\appendix
	
	{\color{black}
		\section{Triadic transfer terms}
		
		The triadic transfer terms $T_{p,q\rightarrow m,n}$ are given by
		\begin{align}
			T_{p,q\rightarrow m,n} &= -\frac{\delta_{m,n}\Rey}{(N_x N_z)^3} \int_0^{L_y} \mathfrak{R} \left( \buhat_{m,n}^* \cdot \left[ \frac{\partial}{\partial y}  (\hat{v}_{m-p,n-q}\buhat_{p,q}+\hat{v}_{m+p,n+q}\buhat_{-p,-q}) \right. \right. \nonumber \\ & \left. \left. +  (imk \hat{u}_{m-p, n-q}+ inl \hat{w}_{m-p,n-q})\buhat_{p,q} +  
			(imk \hat{u}_{m+p, n+q}+ inl \hat{w}_{m+p,n+q})\buhat_{-p,-q} \right]  \right) \,\, \dd y,
		\end{align}
		for $(p,q) \neq (0,0)$, and
		\begin{equation}
			T_{0,0\rightarrow m,n} = -\frac{\delta_{m,n}\Rey}{(N_x N_z)^3} \int_0^{L_y} \mathfrak{R} \left( \buhat_{m,n}^* \cdot \left[ \frac{\partial}{\partial y}  (\hat{v}_{m,n}\buhat_{0,0}) +  (imk \hat{u}_{m, n}+inl \hat{w}_{m, n}) \buhat_{0,0} \right]  \right) \,\, \dd y,
		\end{equation}
		where $\hat{w}_{m,n}$ is the spanwise component of $\buhat_{m,n}$.
	}

	\bibliographystyle{jfm}
	\bibliography{jfm}

@Article{kerczek74,
	author  = {{v}on {K}erczek, C. and Davis, S. H.},
	title   = {Linear stability theory of oscillatory {S}tokes layers},
	journal = {J. Fluid Mech.},
	year    = {1974},
	volume  = {62},
	pages   = {753--773}
}

@Article{Gong2022,
	author  = {Gong, M. and Xiong, C. and Mao, X. and Cheng, L. and Wang, S.-P. and Zhang, A.-M.},
	title   = {Non-modal growth of finite-amplitude disturbances in oscillatory boundary layer},
	journal = {J. Fluid Mech.},
	year    = {2022},
	volume  = {943},
	pages   = {A45}
}

@Article{duguet2013,
  author  = {Duguet, Y. and Monokrousos, A. and Brandt, L. and Henningson, D. S.},
  title   = {Minimal transition thresholds in plane {C}ouette flow},
  journal = {Phys. Fluids},
  year    = {2013},
  volume  = {25},
  pages   = {084103}
}

@Article{sandoval2025,
  author  = {Sandoval, J. and Eaves, T. S.},
  title   = {Edge states and the periodic self-sustaining process in the {S}tokes boundary layer},
  journal = {J. Fluid Mech.},
  year    = {2025},
  volume  = {1022},
  pages   = {A10}
}

@Article{eaves2025,
  author  = {Eaves, T. S.},
  title   = {Nonlinear stability measures of synchronised states in a power-grid model},
  journal = {J. Nonlinear Sci.},
  year    = {2025},
  volume  = {35},
  pages   = {46}
}

@PhDThesis{taylor2008numerical,
  author    = {Taylor, J.R.},
  title     = {Numerical simulations of the stratified oceanic bottom boundary layer},
  year      = {2008},
  school = {University of California, San Diego}
}

@PhDThesis{eavesthesis,
	author    = {Eaves, T. S.},
	title     = {Generalised nonlinear stability of stratified shear flows: adjoint-based optimisation, {K}oopman modes, and reduced models},
	year      = {2016},
	school = {University of Cambridge}
}

@article{eaves2015disruption,
  author  = {Eaves, T.S. and Caulfield, C.P.},
  title   = {Disruption of {SSP/VWI} states by a stable stratification},
  journal = {J. Fluid Mech.},
  year    = {2015},
  volume  = {784},
  pages   = {548--564}
}

@article{waleffe1997self,
  author  = {Waleffe, F.},
  title   = {On a self-sustaining process in shear flows},
  journal = {Phys. Fluids},
  year    = {1997},
  volume  = {9},
  number  = {4},
  pages   = {883--900}
}

@article{hall1991strongly,
  author  = {Hall, P. and Smith, F.T.},
  title   = {On strongly nonlinear vortex/wave interactions in boundary-layer transition},
  journal = {J. Fluid Mech.},
  year    = {1991},
  volume  = {227},
  pages   = {641--666}
}

@article{blennerhassett2008linear,
  author  = {Blennerhassett, P.J. and Bassom, A.P.},
  title   = {On the linear stability of {S}tokes layers},
  journal = {Phil. Trans. R. Soc. A},
  year    = {2008},
  volume  = {366},
  number  = {1876},
  pages   = {2685--2697}
}

@article{blennerhassett2002linear,
  author  = {Blennerhassett, P.J. and Bassom, A.P.},
  title   = {The linear stability of flat {S}tokes layers},
  journal = {J. Fluid Mech.},
  year    = {2002},
  volume  = {464},
  pages   = {393--410}
}

@article{Kerswell2018,
  author  = {Kerswell, R. R.},
  title   = {Nonlinear nonmodal stability theory},
  journal = {Annu. Rev. Fluid Mech.},
  year    = {2018},
  volume  = {50},
  pages   = {319--345}
}

@article{ozdemir2014direct,
  author  = {Ozdemir, C.E. and Hsu, T.-J. and Balachandar, S.},
  title   = {Direct numerical simulations of transition and turbulence in smooth-walled {S}tokes boundary layer},
  journal = {Phys. Fluids},
  year    = {2014},
  volume  = {26},
  number  = {4},
  pages   = {041403}
}

@article{pershin2022optimizing,
  author  = {Pershin, A. and Beaume, C. and Eaves, T.S. and Tobias, S.M.},
  title   = {Optimizing the control of transition to turbulence using a {B}ayesian method},
  journal = {J. Fluid Mech.},
  year    = {2022},
  volume  = {941},
  pages   = {A25}
}

@article{biau2016transient,
  author  = {Biau, D.},
  title   = {Transient growth of perturbations in {S}tokes oscillatory flows},
  journal = {J. Fluid Mech.},
  year    = {2016},
  volume  = {794},
  pages   = {R4}
}

@article{khapko2016edge,
  author  = {Khapko, T. and Kreilos, T. and Schlatter, P. and Duguet, Y. and Eckhardt, B. and Henningson, D.S.},
  title   = {Edge states as mediators of bypass transition in boundary-layer flows},
  journal = {J. Fluid Mech.},
  year    = {2016},
  volume  = {801},
  pages   = {R2}
}

@article{pretty2021,
  author  = {Pretty, A. and Davies, C. and Thomas, C.},
  title   = {Onset of absolutely unstable behaviour in the {S}tokes layer: a {F}loquet approach to the {B}riggs method},
  journal = {J. Fluid Mech.},
  year    = {2021},
  volume  = {928},
  pages   = {A23}
}

@article{ramage2020,
  author  = {Ramage, A. and Davies, C. and Thomas, C. and Togneri, M.},
  title   = {Numerical simulation of the spatio-temporal development of linear disturbances in {S}tokes layers: absolute instability and the effects of high frequency harmonics},
  journal = {Phys. Rev. Fluids},
  year    = {2020},
  volume  = {5},
  pages   = {103901}
}

@article{rabin2014,
	author  = {Rabin, S. M. E. and Caulfield, C. P. and Kerswell, R. R.},
	title   = {Designing a more nonlinearly stable laminar flow via boundary manipulation},
	journal = {J. Fluid Mech.},
	year    = {2014},
	volume  = {738},
	pages   = {R1}
}

@article{rabin2012,
	author  = {Rabin, S. M. E. and Caulfield, C. P. and Kerswell, R. R.},
	title   = {Triggering turbulence efficiently in plane {C}ouette flow},
	journal = {J. Fluid Mech.},
	year    = {2012},
	volume  = {712},
	pages   = {244--272}
}

@article{liftup,
  author  = {Landahl, M.T.},
  title   = {A note on an algebraic instabililty of inviscid parallel shear flows},
  journal = {J. Fluid Mech.},
  year    = {1980},
  volume  = {98},
  pages   = {243--251}
}

@article{orr,
  author  = {Orr, W. M{'}F.},
  title   = {The stability or instability of the steady motions of a perfect liquid and of a viscous liquid. {P}art {I}: {A} perfect liquid},
  journal = {Proc. R. Irish Acad. Sect. A},
  year    = {1907},
  volume  = {27},
  pages   = {9--68}
}

\end{document}